Short Paper

# Applied Computer Vision on 2-Dimensional Lung X-Ray Images for Assisted Medical Diagnosis of Pneumonia

Ralph Joseph S.D. Ligueran
Asian Institute of Computer Studies, Quezon City, Philippines
ligueran170127@gmail.com
(corresponding author)

Manuel Luis C. Delos Santos
Asian Institute of Computer Studies, Quezon City, Philippines
ORCID: 0000-0002-6480-3377

Dr. Ronaldo S. Tinio
Pamantasan ng Lungsod ng Valenzuela, Philippines
ronaldotinio1871@gmail.com

Emmanuel H. Valencia
Asian Institute of Computer Studies, Quezon City, Philippines
emmanuelvalencia008@gmail.com

## Abstract

*Purpose* – This study focuses on the application of a specific subfield of artificial intelligence referred to as computer vision in the analysis of 2-dimensional lung x-ray images for the assisted medical diagnosis of ordinary pneumonia.

*Method* – A convolutional neural network algorithm was implemented in a Python-coded, Flask-based web application that can analyze x-ray images for the detection of ordinary pneumonia. Since convolutional neural network algorithms rely on machine learning for the identification and detection of patterns, a technique referred to as transfer learning was implemented to train the neural network in the identification and detection of patterns within the dataset. Open-source lung x-ray images were used as training data to create a knowledge base that served as the core element of the web application and the experimental design employed a 5-Trial Confirmatory Test for the validation of the web application.

*Results* –The tabulated results of the 5-Trial Confirmatory Test show the calculation of Diagnostic Precision Percentage per Trial, General Diagnostic Precision Percentage, and General Diagnostic Error Percentage while the Confusion Matrix further shows the relationship between the label and the corresponding diagnosis result of the web application on each test images.

*Conclusion* – The successful generation of the h5 knowledge base proved that the CNN algorithm implementation for machine learning can generate patterns based on analyzing open-source datasets. The successful implementation and deployment of the web application to the cloud server proved that such a system can be feasibly deployed in such a platform. The experimental data proved the high precision of the analysis, proved that it can be used in the diagnosis of ordinary pneumonia under the supervision of medical practitioners.

*Recommendations* – In retrospect, the precision of the diagnostic results could be enhanced further by utilizing a much larger training dataset for the machine learning phase. Since machine learning algorithms rely on the probability theory concept of the Law of Large Numbers, a large, high-quality dataset is crucial in yielding high-precision results. A more balanced training dataset is also necessary to avoid any kind of statistical bias during machine learning. A more balanced dataset could negate the use of data augmentation functionalities, thereby improving the efficiency and speed of the machine learning phase.

*Practical Implications* – The developed web application can be used by medical practitioners in A.I.-assisted diagnosis of ordinary pneumonia, and by researchers in the fields of computer science and bioinformatics.

---

## INTRODUCTION

The advancements of computer vision technology have led to several significant milestones in the field of artificial intelligence. Computer vision technologies have been applied to various platforms, such as social networking sites, closed-circuit television camera networks, and robotics among others. Recently, due to the growing interest in the development of A.I.-assisted medical diagnostic technologies and healthcare/biomedical informatics, computer vision is slowly but gradually being utilized in the diagnosis of diseases (Hamet & Tremblay, 2017).

Medical images, such as x-ray images as shown in Figure 1 and CT scans, are highly useful in the diagnosis of certain diseases. Using computer vision, the detection and analysis of these diseases can theoretically yield higher accuracy and reliability in medical diagnosis.

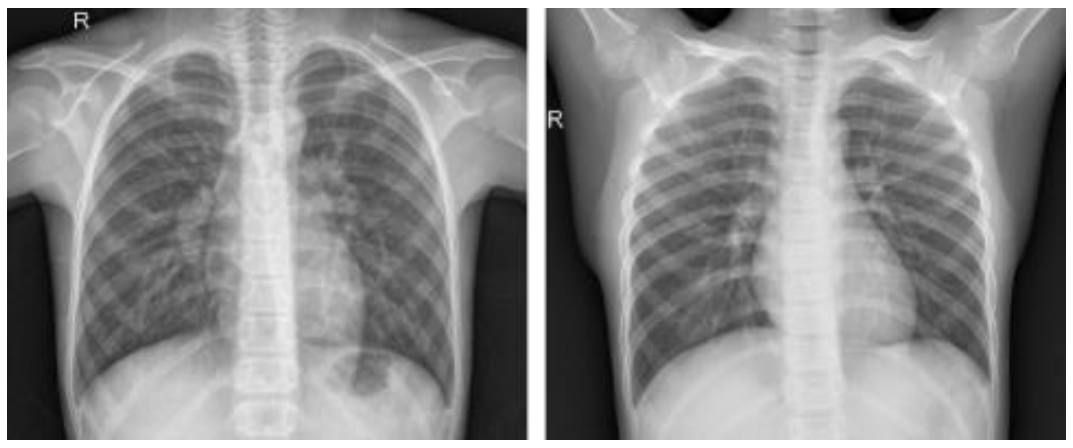

*Figure 1*. Dimensional Lung X-Ray Images (Dincer, 2020).

There exist various algorithms for the implementation of computer vision technologies. One of them is the Convolutional Neural Network (CNN). A convolutional neural network is a type of artificial neural network. Artificial neural networks are either hardware or software implementations that imitate the structures and operation of organic neural networks that exist in the human brain. An organic neural network is composed of neurons and their corresponding dendrite-axon connections with one another.

A convolutional neural network, therefore, as observed in Figure 2 is a deep neural network based on the mathematical operation of convolution that relies on machine learning to detect patterns within datasets. For the convolutional neural network to "learn", training datasets must be fed into the neural network during the machine learning phase (Khan & Al-Habsi, 2019).

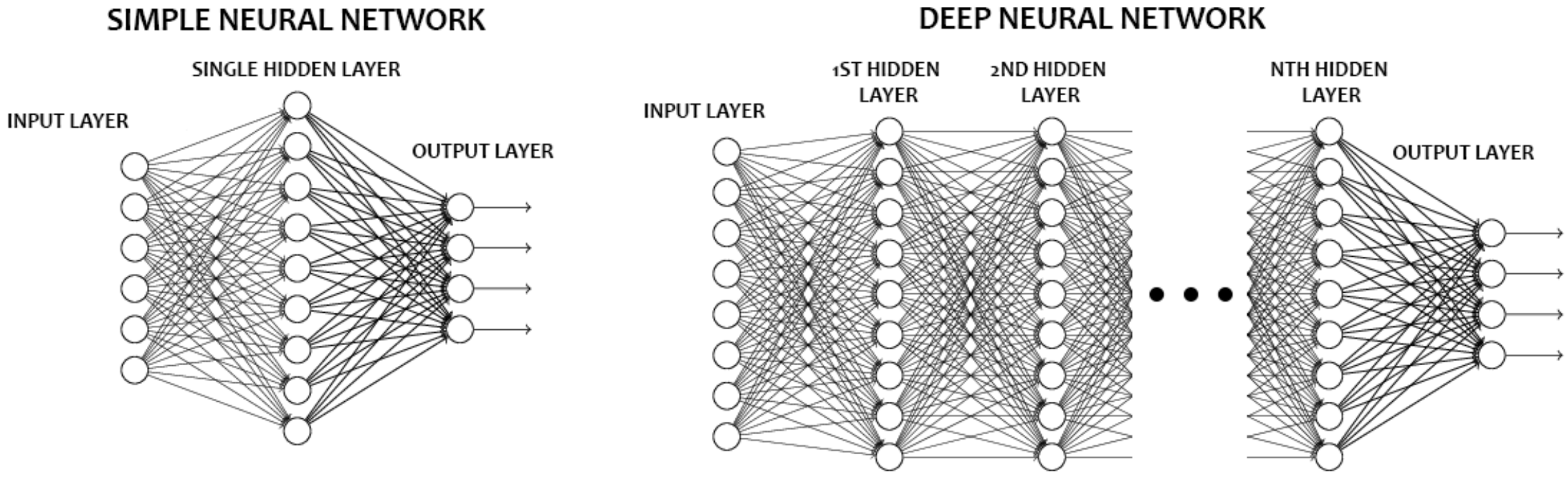

*Figure 2*. Comparison of Simple and Deep Neural Networks.

The primary concern in the development of this kind of A.I.-assisted medical diagnostic technology is to prevent cases of medical misdiagnosis. Medical misdiagnoses are cases where medical practitioners unintentionally commit errors in the diagnosis of diseases on patients. False negatives or false positive misdiagnosis can occur due to various factors. Human error can also be a significant factor in the causation of medical misdiagnosis. Such cases of medical misdiagnosis of pneumonia can lead to life-threatening conditions.

Analysis of medical images is also essential in the discovery of new types of diseases. At the time of the writing of this research paper, the COVID-19 pandemic is still ravaging various nations around the world. Before the implementation of nasal swab testing for the detection of the presence of the SARS-Cov-2 virus in patients, signs of the COVID-19 disease were first detected through lung x-ray images. Due to the novelty of the disease, the x-ray results were misdiagnosed as simply ordinary pneumonia. However, as more COVID-19 x-ray images were gathered and compared with pre-existing x-ray images of pneumonia, patterns were later revealed. This case exemplifies the importance of medical imaging in the diagnosis of diseases (Cellina, et al., 2020). Due to the inherently broad scope of artificial intelligence in medicine, this research only focused on the detection of ordinary pneumonia.

Medical imaging can therefore be improved to produce more accurate results through the help of artificially intelligent tools. Artificial intelligence should be regarded as just a tool for solving various problems. Human ingenuity combined with the strengths of artificial intelligence could solve the majority of the world's problems (Ito, 2016).

The objectives of the study aim for the following:

1. The implementation of the Convolutional Network Algorithm and Transfer Learning Technique, and the utilization of open-source datasets for knowledge base generation through machine learning.
2. The development of a Flask-based web application using the generated knowledge base.
3. The verification of the precision of the developed web application using experimentation and statistical techniques.

## LITERATURE REVIEW

### *A. Convolutional Neural Networks and Machine Learning for Computer Vision*

The concepts of supervised, unsupervised, and semi-supervised machine learning and various new and emerging techniques on how to implement these concepts in computational machines will advance further as new and more algorithms are developed by Khan and Al-Habsi (2019). Deep learning will be the mainstream technology in medical imaging in the next few decades (Suzuki, 2017). The development of the VGGNet algorithm for computer vision and its variant algorithms with accuracy and precision rating results ranging from 70% to 99% on various testing metrics for the algorithms (Simonyan & Zisserman, 2015).

### *B. Application of Computer Vision on 2-Dimensional Lung X-ray Images for the Detection of Ordinary Pneumonia*

According to research articles regarding the application of computer vision algorithms on 2-D X-ray images for the detection of ordinary pneumonia authored by Ayan et al. (2021), Hashmi et al. (2020), Kundu et al. (2021), and Yue et al. (2020), the typical precision rating ranges from 90% to 98%.

These research articles were indexed in PubMed - National Center for Biotechnology Information. NCBI is a United States government institution that aims to advance scientific and medical research by providing access to various biomedical and genomic information.

### *C. Web Frameworks*

The issue of the lack of analytic tools for the evaluation of the performance of various Flask-based web applications had led to its development (Vogel et al., 2017). Web developers should be aware of the pros and cons for the efficient deployment of web applications (Curie, et al., 2019).

## METHODOLOGY

### *A. The Mathematics of Convolution – A Functional Analysis Perspective*

In functional analysis, convolution is defined as the operation between two functions *f* and *g* that expresses a third function that indicates how the shape of the graph of one is modified by the other (Kutateladze, 1996). If *f* and *g* are discrete functions, then *f* * *g* is the convolution of *f* and *g* as defined in Equation 1:

$$(\boldsymbol{f} * \boldsymbol{g})(\boldsymbol{t}) \coloneqq \int_{\infty}^{\infty} \boldsymbol{f}(\boldsymbol{r})\boldsymbol{g}(\boldsymbol{t}-\boldsymbol{r})\, \boldsymbol{dr}. \qquad \textit{Equation 1}$$

B. The Mathematics of Convolution – A Matrix Theory Perspective

In matrix theory, convolution is defined as the recursive dot product of two matrices. Instead of generalized functions, the operation is done with matrices, A and B, where matrix B is of smaller size than matrix A. Matrix B is referred to as the kernel matrix. The kernel matrix contains the pattern that can be found in matrix A. After a recursive dot product of matrix A to B, the resultant matrix is obtained. If the resultant matrix is the same or similar to the kernel matrix, then it can be concluded that the pattern stored in the kernel matrix exists in the input matrix (Hunt, 1971). If *A* and *B* are matrices, then *A* * *B* is the convolution of *A* and *B* defined in Equation 2:

$$\begin{bmatrix} x_{11} & x_{12} & \cdots & x_{1n} \\ x_{21} & x_{22} & \cdots & x_{2n} \\ \vdots & \vdots & \ddots & \vdots \\ x_{m1} & x_{m2} & \cdots & x_{mn} \end{bmatrix} \begin{bmatrix} y_{11} & y_{12} & \cdots & y_{1n} \\ y_{21} & y_{22} & \cdots & y_{2n} \\ \vdots & \vdots & \ddots & \vdots \\ y_{m1} & y_{m2} & \cdots & y_{mn} \end{bmatrix}$$

$$= \sum_{i=0}^{m-1} \sum_{j=0}^{n-1} x_{(m-i)(n-j)} y_{(1+i)(1+j)} \qquad \text{Equation 2}$$

### C. *Convolutional Neural Network in Computer Vision*

As presented in Figure 3, a convolutional neural network is used primarily on datasets that can be represented as matrices. This means CNN is highly utilized in the field of computer vision, where 2-dimensional images can be represented as matrices. Representation of images into matrices can be achieved using two techniques: either every pixel is assigned a numeric value, converting the entire image into a matrix, or a separate algorithm is used to partition the image into arbitrary sizes, with each grid being assigned a numeric value (Khan & Al-Habsi, 2019).

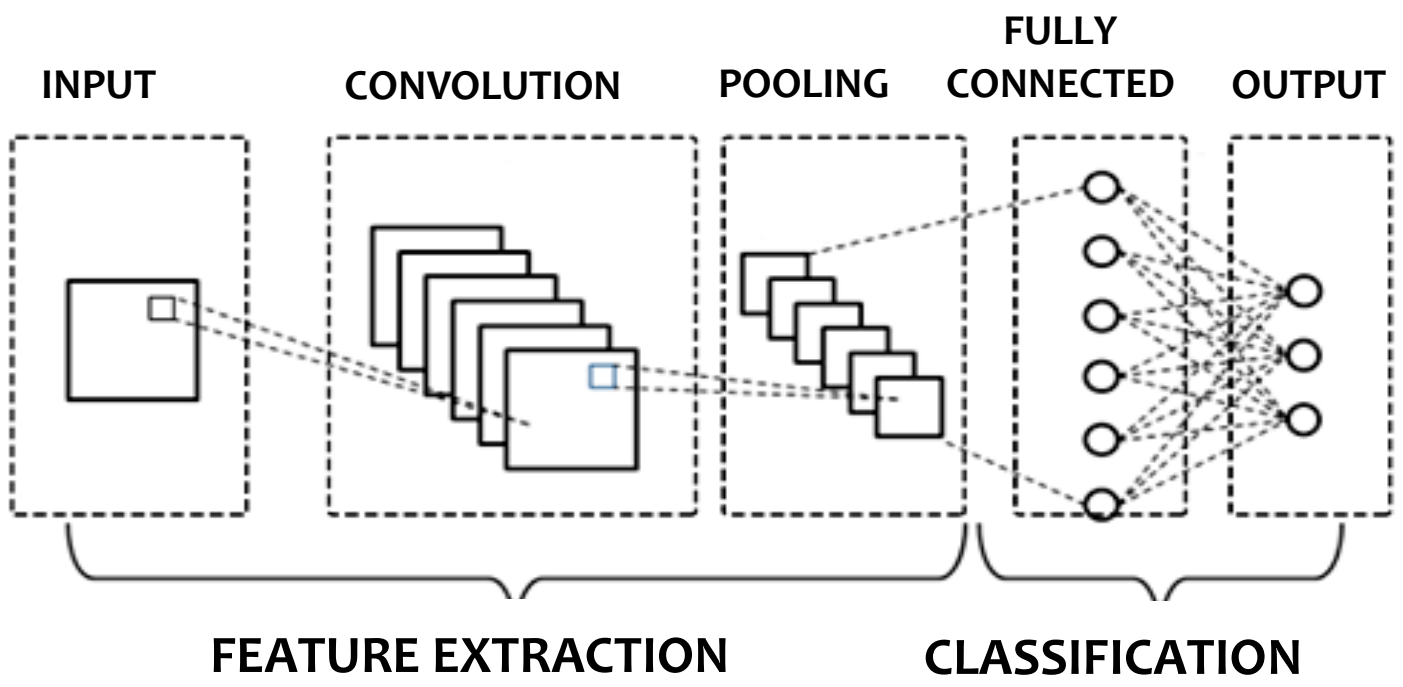

*Figure* 3. General Concept of a Convolutional Neural Network Algorithm.

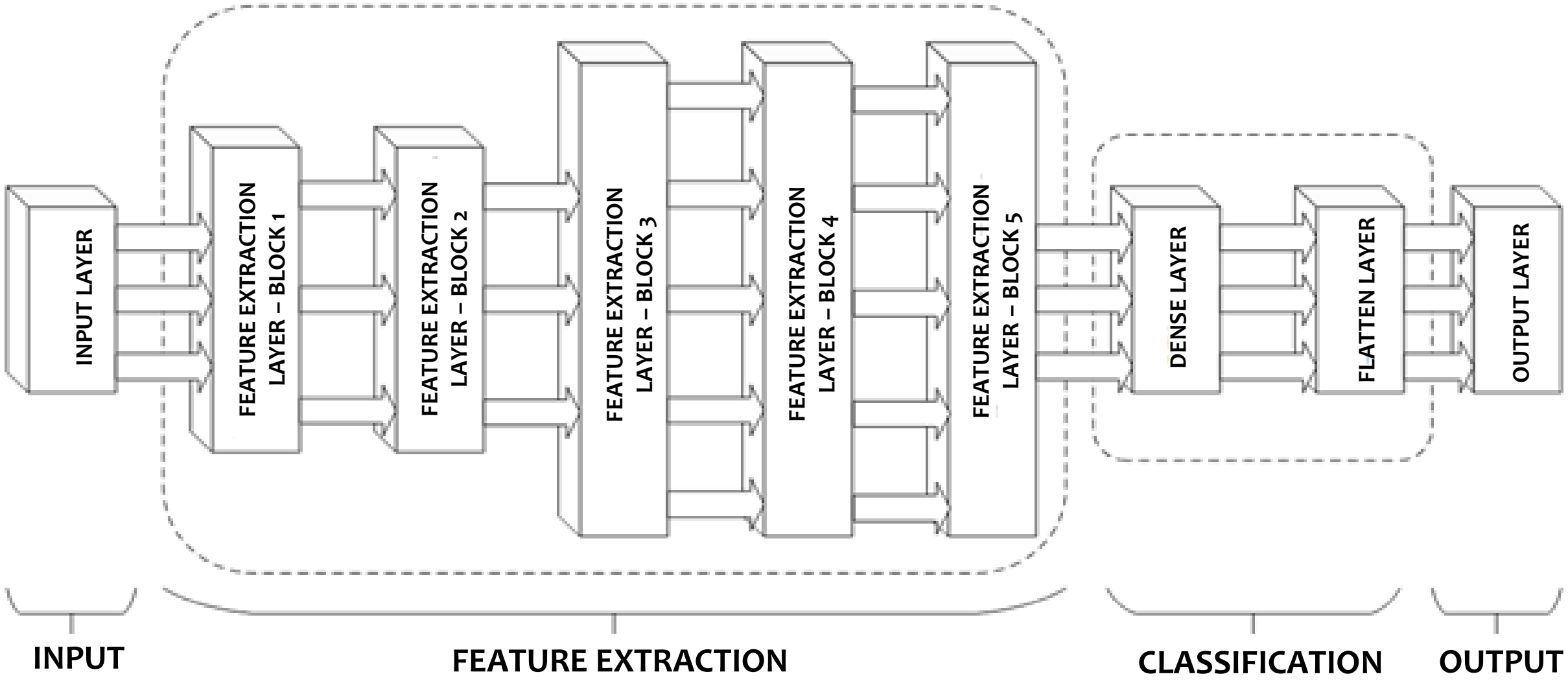

*Figure* 4. Block Diagram of the VGG-16 Convolutional Network Algorithm.

The VGG-16 algorithm, as depicted in Figure 4 was developed by Simonyan and Zisserman (2015) is primarily composed of five blocks of feature extraction layers, a dense layer, and a flattened layer. Each block of the feature extraction layer is composed of Cov2D and MaxPooling2D stages as illustrated in Figure 5. Cov2D is a computational function that can execute a convolutional operation on images. MaxPooling2D is a computational function that minimizes the size of the matrix by getting the highest value within a partition. For feature extraction layers of blocks 1 and 2, the matrices will pass through 2 stages of Cov2D, and a single stage of MaxPooling2D. For blocks 3, 4, and 5, the matrices will pass through three stages of Cov2D and a single stage of MaxPooling2D.

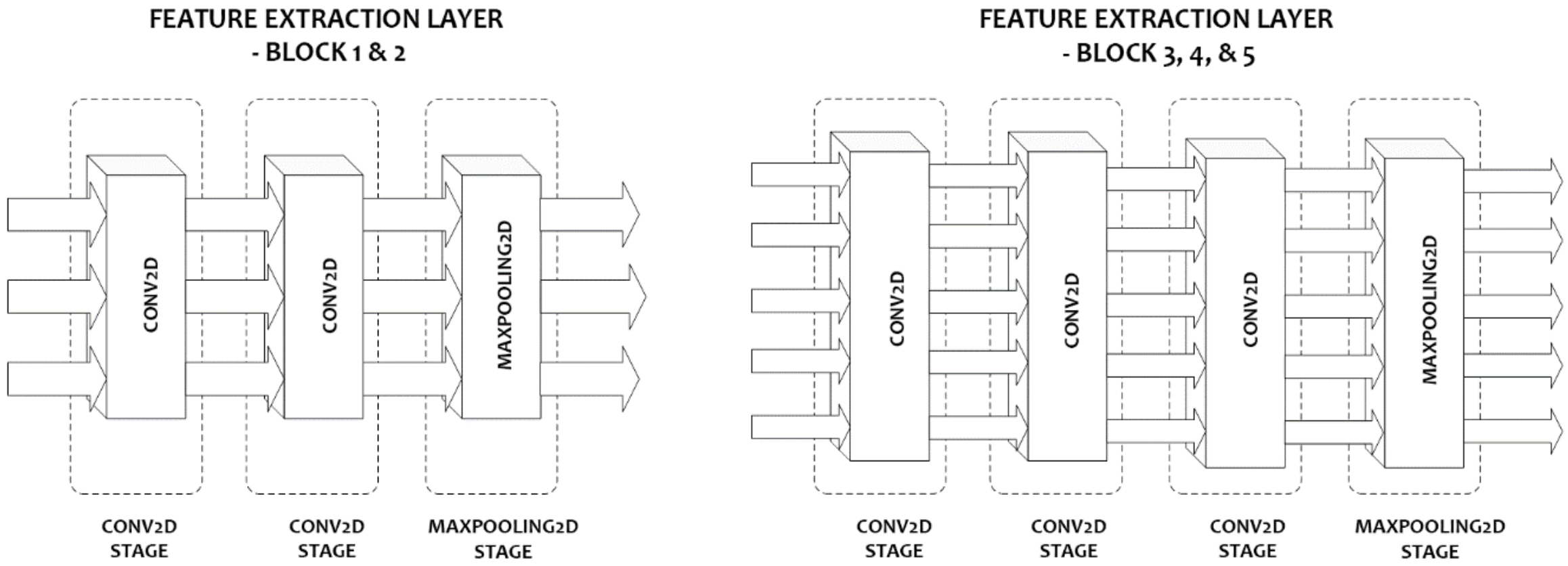

*Figure* 5. Feature Extraction Layers with their various Conv2D - Maxpooling2D stage configurations.

### D. *Transfer Learning Technique*

Transfer learning as revealed in Figure 6 is an alternative technique in machine learning that combines supervised and unsupervised machine learning. Supervised machine learning is when training data is analyzed by the application, with the resulting analysis data applied during operational phases. Unsupervised machine learning is when the application acquires analysis data during the operational phase itself. Transfer learning allows the application to make use of the strengths of both supervised and unsupervised machine learning by allowing the application to acquire analysis data and store it on a pre-existing knowledge base (Suzuki, 2017).

In this study, a machine learning model file with an h5 file extension contains all the data relevant to the analysis of the image. Kernel matrices containing the pattern to be detected are stored in the h5 file.

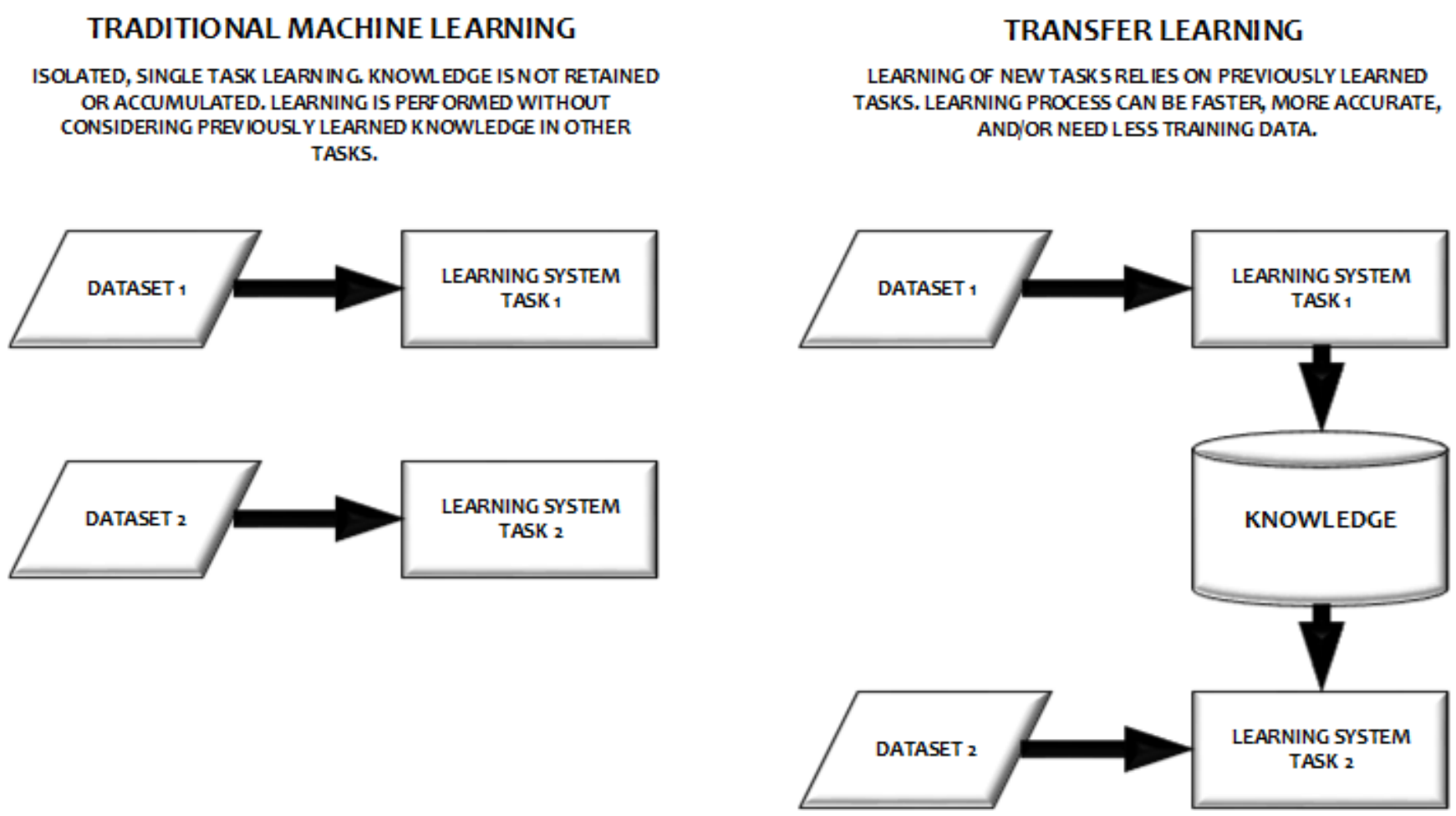

*Figure 6*. Comparison of traditional machine learning and transfer learning.

## *E. Training Datasets*

Training datasets were obtained from Kaggle, an open-source research community owned by Google LLC, which specializes in data analytics and machine learning research. Kaggle's objective is to provide free and high-quality open-source datasets from various fields of study to further research in data analytics and machine learning.

The dataset used is entitled "Chest X-ray Images for Pneumonia Detection with Deep Learning" published by Tolga Dincer (2020), a data scientist from Connecticut, United States. The dataset was selected due to its generally good reviews, recent publication date, and high Kaggle usability rating of 8.1.

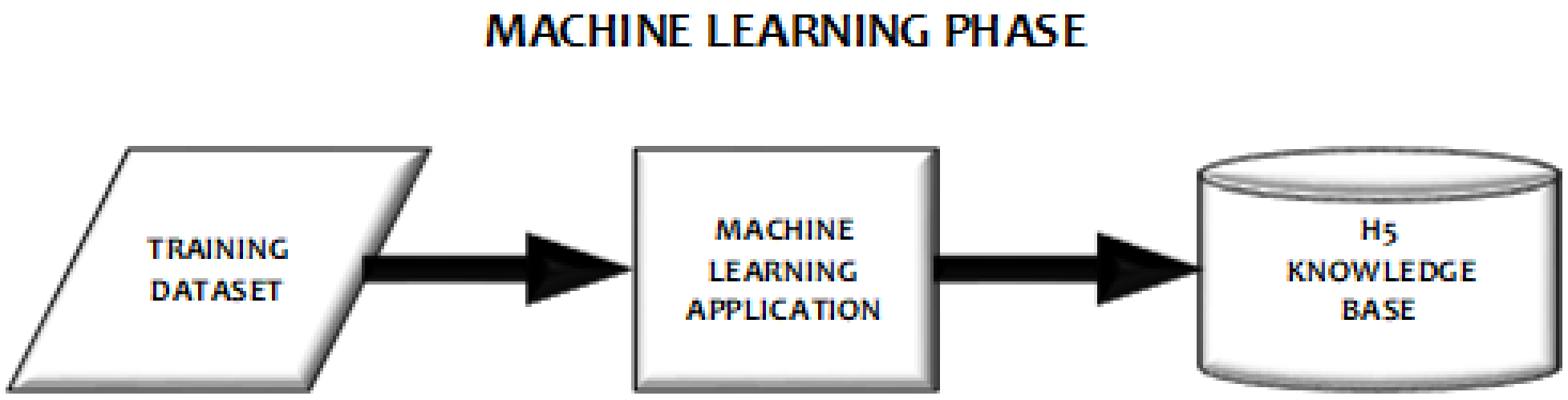

*Figure 7*. Machine Learning Phase Diagram for Knowledge Base Generation

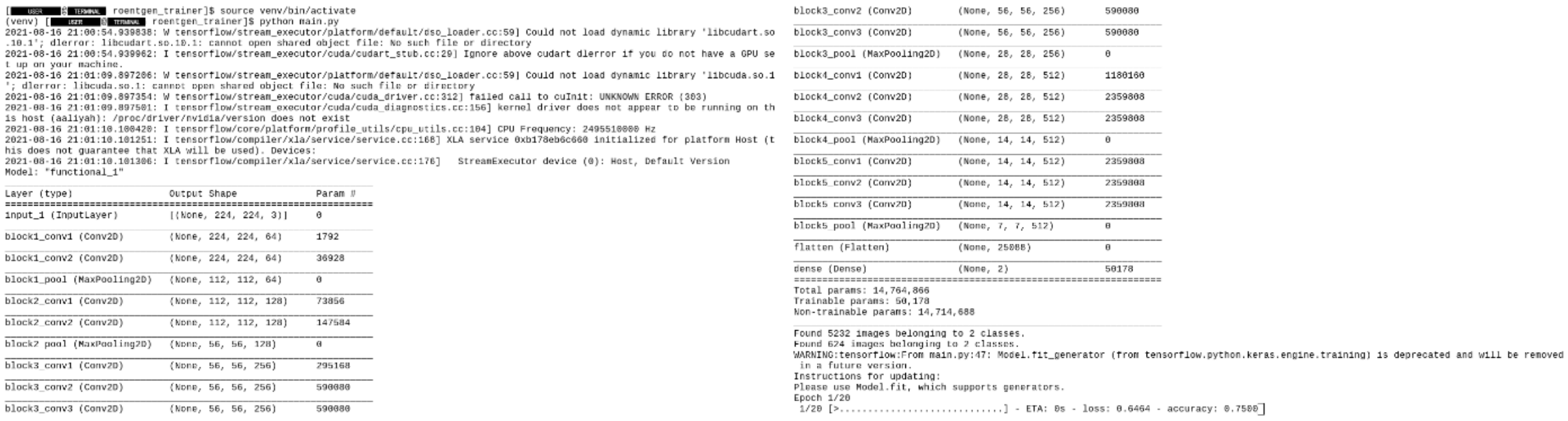

*Figure 8*. VGG-16 machine learning application output in local machine's bash console.

The machine learning phase and the sample VGG-16 application output are demonstrated in Figure 7 and Figure 8.

### *F. Implementation Using Flask*

To implement the image analyzer and utilize the generated h5 knowledge base file for the detection of ordinary pneumonia, a Python application was developed around a Flask framework for web deployment. PythonAnywhere, a Python-compatible cloud server was used to host the web application, including the virtual machine and cloud storage. As explained in Figure 9, the web application system architecture that the proponents have developed follows the standard web application architecture. The front-end is composed of HTML & CSS-coded web pages. The back-end is created using the Python-coded Flask framework, and the storage is where the h5 file and image storage directory are located (Curie et al., 2019; Vogel et al., 2017).

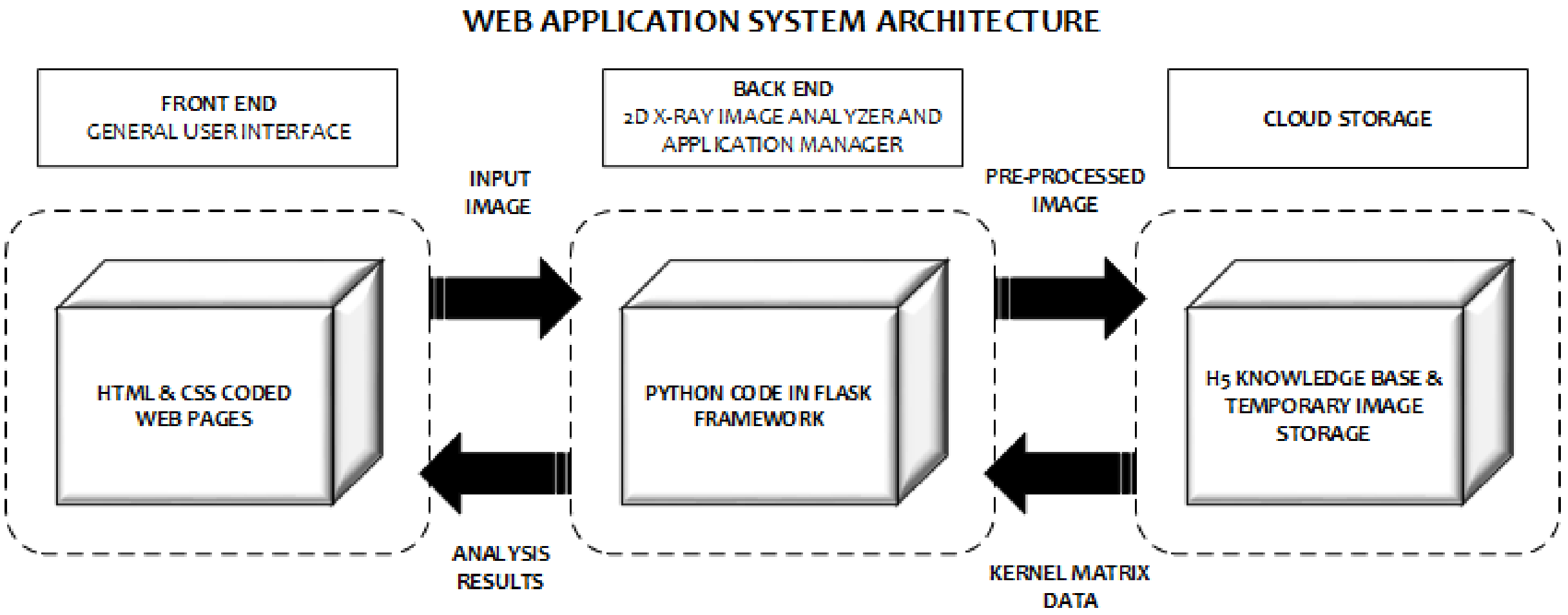

*Figure 9*. Web Application System Architecture

The web application was publicly accessible and hosted in the domain www.roentgenanalysis.pythonanywhere.com (currently down due to unpaid subscription in PythonAnywhere). All web pages of the web application as displayed in Figure 10 are accessible to the users. The list of web pages are:

1. Home Page – shows the general overview of the web application and image uploader. The diagnosis results will be rendered here or can be exported as PDF.
2. Transfer Learning page – contains the overview of the machine learning process, as well as the graph outputs.
3. Extended Intelligence page – explains the concept of Extended Intelligence.
4. Credit page - contains a short profile of the developers, research adviser, and the abstract of the research.

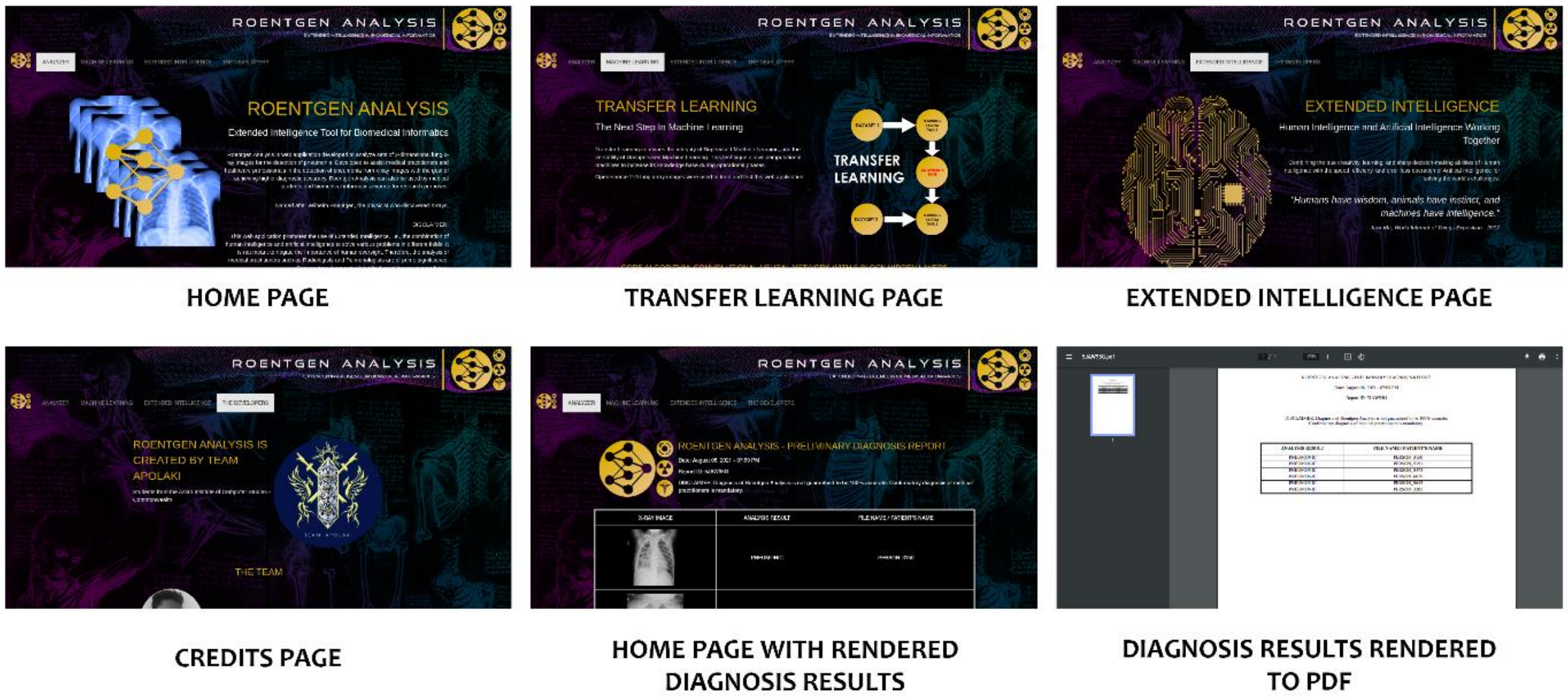

*Figure 10*. Web Application User Interface

## G. *Quantitative Approach, Experimental Design, and Statistical Methods for Precision Validation*

To test the precision and effectiveness of the developed web application, a confirmatory test was conducted using a quantitative approach and statistical methods. The experimental design was a 5-trial confirmatory test. In this approach, a confirmatory test was conducted five times using five random and independently selected test datasets. For each trial, a sample of one-hundred 2-dimensional lung x-ray images was gathered from the open-source dataset. The selection process was conducted using a statistical approach known as Simple Random Sampling without Replacement (SRSWOR).

In SRSWOR, a sample was selected from the population randomly with no duplicates. This sample served as the test dataset. Fifty images that were labeled as pneumonic were selected using SRSWOR, while another fifty that were labeled as non-pneumonic were also selected using SRSWOR. These fifty pneumonic and non-pneumonic 2-D lung x-ray images composed the test dataset of size 100 per trial. The 1:1 ratio of pneumonic and non-pneumonic images ensured less statistical bias. These test datasets were then analyzed using the web application. Since SRSWOR was used in sampling, all five 100-sample test datasets have unique elements. There is no duplication of images for all sample sets.

To verify the accuracy of the web application, the resulting preliminary diagnosis should match the label of each x-ray image. To do this, three statistical measures were formulated for this specific experimental design, the Diagnosis Precision Percentage per Trial (DPP-Ti), General Diagnostic Precision Percentage (GDPP), and the General Diagnostic Error Percentage (GDEP). The Diagnostic Precision Percentage per Trial is defined as the arithmetic mean of the matched diagnosis in a specific trial, multiplied by 100%. To compute this, the number of matched diagnoses was tallied, and the number is divided by the sample size, which is 100. This arithmetic mean is then multiplied to 100%. The General Diagnostic Precision Percentage is defined as the arithmetic mean of all five Diagnostic Precision Percentages per Trial. This is computed by summing all five DPP-Ti and dividing it by 5. The General Diagnosis Error Percentage is defined as the error rating and is computed by subtracting the GDPP from 100% (Peck et al., 2008).

The experimental procedure is as follows:
Step 1: Upload the test dataset into the web application.
Step 2: Run web application diagnosis.
Step 3: Gather diagnosis results.
Step 4: Verify diagnosis results with test datasets and tally all matched and unmatched results.
Step 5: Compute for the Diagnosis Precision Percentage per Trial (DPP-Ti).
Step 6: Repeat Steps 1 to 5 for all 5 100-sample test datasets.
Step 7: Compute for the General Diagnostic Precision Percentage (GDPP) and General Diagnostic Error Percentage (GDEP).

### *H. Statistical Formulas for Precision Validation*

Arithmetic Mean *($\bar{x}$)* with sample size n as defined in Equation 3:

$$\dot{\bar{\mathrm{X}}} = \frac{\sum_{i=1}^{n} xi}{\mathrm{n}}$$ *Equation 3*

Diagnostic Precision Percentage per Trial (DPP-Ti) as defined in Equation 4:

$$DPP - Ti = \frac{\text{Total Number of Matched Diagnosis}}{100} \text{x } 100\%$$ *Equation 4*

General Diagnostic Precision Percentage (GDPP) as defined in Equation 5:

$$\mathrm{GD\dot{P}P} = \frac{\sum_{\mathrm{i}=1}^{5} \mathrm{DPP - Ti}}{5}$$ *Equation 5*

General Diagnostic Error Percentage as defined in Equation 6:

$$GDEP = 100\% - GDPP$$ *Equation 6*

# RESULTS

## *A. Machine Learning and Knowledge Base Generation*

The screenshot in Figure 11 below showed the successful generation of the h5 knowledge base from the VGG-16 machine learning application.

```
$ cd PycharmProjects
$ cd roentgen_trainer
$ ls -la
total 58304
drwxr-xr-x  5    USER    wheel     4096 Feb 12 15:34 .
drwxr-xr-x 26    USER    wheel     4096 Dec 16 19:24 ..
drwxr-xr-x  5    USER    wheel     4096 Apr  1  2021 chest_xray
-rw-r--r--  1    USER    wheel    54237 Apr  2  2021 epochs.png
drwxr-xr-x  3    USER    wheel     4096 Feb 12 15:40 .idea
-rw-r--r--  1    USER    wheel     2953 Feb 12 15:34 main.py
-rw-r--r--  1    USER    wheel 59540320 Apr  2  2021 model.h5
-rw-r--r--  1    USER    wheel    73261 Aug 16  2021 model.png
-rw-r--r--  1    USER    wheel      792 Mar 15  2021 req.txt
drwxr-xr-x  5    USER    wheel     4096 Mar 24  2021 venv
$ stat model.h5
  File: model.h5
  Size: 59540320        Blocks: 116296     IO Block: 4096   regular file
Device: 8,4     Inode: 9984864     Links: 1
Access: (0644/-rw-r--r--)  Uid: ( 1000/ialzayani)   Gid: (  998/   wheel)
Access: 2022-02-12 15:32:21.397615603 +0800
Modify: 2021-04-02 03:32:37.618554330 +0800
Change: 2021-04-02 03:32:37.618554330 +0800
 Birth: 2021-04-02 03:32:37.501887659 +0800
$ 
```

*Figure 11*. Local machine bash console showing the generated h5 file.

## *B. Cloud Deployment*

The screenshot in Figure 12 below showed the PythonAnywhere bash console, proving the successful deployment of the web application into the cloud server.

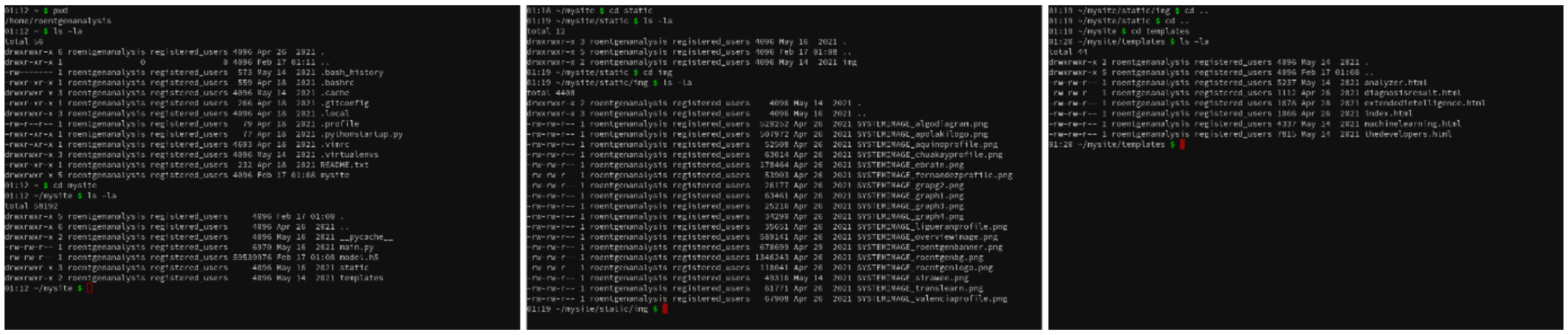

*Figure 12*. PythonAnywhere bash console, showing the project files stored and deployed in the cloud server.

## *C. Experimental Results*

Tabulated result in Table 1 of the 5-Trial Confirmatory Test, with the calculated Diagnostic Precision Percentage per Trial, General Diagnostic Precision Percentage, and General Diagnostic Error Percentage and the Confusion Matrix in Table 2.

Table 1. Result of the 5-Trial Confirmatory Test

| | |
|---|---|
| **Trial 1** | |
| Matched Diagnosis | Tally Result: 92 |
| Unmatched Diagnosis | Tally Result: 8 |
| Diagnosis Precision Percentage | 92 % |
| **Trial 2** | |
| Matched Diagnosis | Tally Result: 93 |
| Unmatched Diagnosis | Tally Result: 7 |
| Diagnosis Precision Percentage | 93 % |
| **Trial 3** | |
| Matched Diagnosis | Tally Result: 90 |
| Unmatched Diagnosis | Tally Result: 10 |
| Diagnosis Precision Percentage | 90 % |
| **Trial 4** | |
| Matched Diagnosis | Tally Result: 91 |
| Unmatched Diagnosis | Tally Result: 9 |
| Diagnosis Precision Percentage | 91 % |
| **Trial 5** | |
| Matched Diagnosis | Tally Result: 90 |
| Unmatched Diagnosis | Tally Result: 10 |
| Diagnosis Precision Percentage | 90 % |
| **Final Result** | |
| General Diagnosis Precision Percentage: | **91.2** % |
| General Diagnosis Error Percentage: | **8.8** % |

Table 2. Confusion Matrix

| | Labeled As "Pneumonic" | Labeled As "Not Pneumonic" |
|---|---|---|
| Diagnosed as "Pneumonic" | *TRUE POSITIVE*<br><br>Tally 1:50<br>Tally 2: 50<br>Tally 3: 50<br>Tally 4: 50<br>Tally 5: 50 | *FALSE-POSITIVE*<br><br>Tally 1: 8<br>Tally 2: 7<br>Tally 3: 10<br>Tally 4: 9<br>Tally 5: 10 |
| Diagnosed as "Not Pneumonic" | *FALSE NEGATIVE*<br><br>Tally 1: 0<br>Tally 2: 0<br>Tally 3: 0<br>Tally 4: 0<br>Tally 5: 0 | *TRUE NEGATIVE*<br><br>Tally 1: 42<br>Tally 2: 43<br>Tally 3: 40<br>Tally 4: 41<br>Tally 5: 40 |

## DISCUSSION

### *A. Statistical Interpretation and Justification*

A General Diagnostic Precision Percentage of 91.2% and a General Diagnosis Error Percentage of 8.8 % were computed from the 5-Trial Confirmatory Test. This means that the developed web application is precise with its diagnosis 91.2% of the time, and can commit errors in its diagnosis for 8.8% of the time. This result is consistent with the research conducted by Ayan et al. (2021), Hashmi et al. (2020), Kundu et al. (2021), and Yue et al. (2020), concerning the implementation of computer vision algorithms for the detection of ordinary pneumonia, where precision ratings range from 90% to 98%. Since their research articles were indexed by PubMed – National Center for Biotechnology Information (NCBI), a reputable and credible research institution in the United States, we can confidently attest to the factuality of their results.

### *B. Implication and Significance*

The confusion matrix illustrates that the web application commits no errors in the diagnosis of pneumonia on pneumonic-labeled images, but commits errors when diagnosing non-pneumonic images. The error is that certain non-pneumonic images were diagnosed as pneumonia, which is a false positive error. There are no false-negative errors found. This means for 8.8 % of the time, errors in the diagnosis will be false positive. This result is significant since false-negative errors have generally more severe consequences than false-positive errors.

## CONCLUSIONS AND RECOMMENDATIONS

The authors, therefore, concluded that the developed web application is precise for 91.2% of the time. A GDEP value of 8.8% tells that an error in the diagnosis can happen 8.8% of the time. The confusion matrix of the experimental result illustrates that only false positive errors are possible with no false-negative errors found. This error is reasonable since false positive errors (non-pneumonic but diagnosed as pneumonic) have less serious consequences than false-negative errors (pneumonic but diagnosed as non-pneumonic).

The error can be attributed to the inherent limitation of the VGG-16 algorithm, as well as the quality of the training and test datasets. Since transfer learning relies heavily on the training datasets, the quality of the analysis is linked directly to the quality of the dataset. High-quality datasets in large quantities are essential in yielding high-precision diagnoses. Small errors within the test dataset can manifest in the generated knowledge base and analysis result. Various iterations of the CNN algorithm will yield various precision ratings.

The successful generation of the h5 knowledge base proved that the CNN algorithm implementation for machine learning can generate patterns based on analyzing open-source datasets. The successful implementation and deployment of the web application to the cloud server proved that such a system can be feasibly deployed in such a platform. The experimental data proved the high precision of the analysis, proving that it can be used in the diagnosis of ordinary pneumonia under the supervision of medical practitioners.

In retrospect, the precision of the diagnostic results could be enhanced further by utilizing a much larger training dataset for the machine learning phase. Since machine learning algorithms rely on the probability theory concept of the Law of Large Numbers, a large, high-quality dataset is crucial in yielding high-precision results. A more balanced training dataset is also necessary to avoid any kind of statistical bias during machine learning. A more balanced dataset could negate the use of data augmentation functionalities, thereby improving the efficiency and speed of the machine learning phase.

## IMPLICATIONS

The developed web application can be used by medical practitioners in A.I.-assisted diagnosis of ordinary pneumonia, and by researchers in the fields of computer science and bioinformatics.

## ACKNOWLEDGEMENT

The researchers would like to extend their sincerest gratitude to the authors of various works of literature used in this study and all the people who have made the development and publication of this paper amidst the pandemic COVID-19 possible. Their contributions to the creation of new knowledge not only led to the advancement of their fields of study but the advancement of humanity in general.